\documentclass{article}
\PassOptionsToPackage{numbers, compress}{natbib}
\usepackage{natbib}
\usepackage[final]{neurips_2022}
\usepackage[utf8]{inputenc} 
\usepackage[T1]{fontenc}    
\usepackage{url}            
\usepackage{booktabs}       
\usepackage{amsfonts}       
\usepackage{nicefrac}       
\usepackage{microtype}      
\usepackage{xcolor}         
\usepackage{caption}
\usepackage{subcaption}
\usepackage{array, longtable}
\usepackage{amssymb}
\usepackage{graphicx}
\usepackage{pifont}
\usepackage{epsfig}   
\usepackage{amsmath}
\usepackage{mathptmx}
\usepackage{txfonts}
\usepackage[T1]{fontenc}
\usepackage{ae,aecompl}
\usepackage{amssymb}	
\usepackage{scalerel}
\usepackage{longtable}
\usepackage{titlesec}
\usepackage{graphicx,epsfig}
\usepackage{psfrag}
\usepackage{inputenc}
\usepackage{xcolor}
\usepackage{hyperref}

\title{Time evolution of the mutual information between disjoint regions in the Universe}

\author{Biswajit Pandey \\
  Department of Physics, Visva-Bharati University, Santiniketan 731235, India\\
  \texttt{biswap@visva-bharati.ac.in} 
}   

\begin{document}

\maketitle

\begin{abstract}
 We study the time evolution of the mutual information between the
 mass distributions in spatially separated but casually connected
 regions in an expanding universe. The evolution of the mutual
 information is primarily determined by the configuration entropy rate
 which depends on the dynamics of the expansion and the growth of the
 density perturbations. The joint entropy between the distributions
 from the two regions plays a negligible role in such evolution. The
 mutual information decreases with time in a matter dominated Universe
 whereas it stays constant in a $\Lambda$-dominated Universe. The
 $\Lambda$CDM model and some other models of dark energy predict a
 minimum in the mutual information beyond which the dark energy
 dominates the dynamics of the Universe. The mutual information may
 have deeper connections to the dark energy and the accelerated
 expansion of the universe.
\end{abstract}

\section{Introduction}

Entropy plays a key role in understanding a wide range of phenomena in
science. It plays an important role in deciding the evolution of the
universe. The Universe is regarded as a dynamical system in
Cosmology. The dynamics of the expansion of the universe is described
by the Friedmann equation which is based on the Einstein's equation of
General relativity and the cosmological principle. The cosmological
principle assumes that the universe is statistically homogeneous and
isotropic on sufficiently large scales. The validity of this
assumption is crucial to our understanding of the modern cosmology. A
large number of studies have been carried out to verify the
cosmological principle. Studies based on various cosmological
observations find that the universe is statistically homogeneous and
isotropic on scales somewhere beyond $70-150$ Mpc
\citep{martinez94,borgani95,guzzo97,cappi,bharad99,pan2000,yadav,hogg,prakash,scrim,nadathur,pandeysarkar15,pandeysarkar16}.
The universe is highly inhomogeneous and anisotropic on smaller scales
due to the presence of a clear hierarchy of structures starting from
planets, stars, galaxies, groups, clusters to superclusters. All these
structures in the present universe are believed to emerge from the
growth of the primordial density fluctuations seeded in the early
universe. The observed CMBR temperature fluctuations of $\frac{\Delta
  T}{T} \sim 10^{-5}$ at redshift $z\sim 1100$ suggests that the
universe was highly homogeneous and isotropic in the past. These tiny
fluctuations are amplified by the gravitational instability to produce
structures over a wide range of length scales.

Our universe is known to be expanding. Recent observations suggest
that the universe is currently undergoing an accelerated expansion
\citep{riess98, perlmutter99}.  Understanding the present accelerated
expansion of the universe is a major unsolved problem in
cosmology. The dynamics of the expansion affect the growth of
inhomogeneities in the universe. Conversely, the inhomogeneities may
also play an important role in the observed acceleration through their
effect on the large-scale dynamics of the universe \citep{buchert97,
  schwarz, kolb06, buchert08, ellis}. \cite{pavon14} suggests that the
observed acceleration of the universe is consistent with the second
law of thermodynamics and the entropy of the universe in the
$\Lambda$CDM model tends to a finite maximum. Interestingly, the
alternative models such as non-singular bouncing Universes, modified
gravity theories and phantom fields do not lead to a state of maximum
entropy \citep{radicella12, mimoso13, ferreira16}. The continuous
dissipation of the information entropy of the matter distribution due
to the gravitational instability may also drive the accelerated
expansion of the universe \citep{pandey17, pandey19}. These studies
suggest that the large-scale inhomogeneity and the maximum entropy
production principle (MEPP) \citep{martyushev06, martyushev13} may
have important roles in the observed acceleration of the universe.

Numerous works in the literature point to the existence of very
large-scale structures in the universe. The `Sloan Great Wall' in the
nearby universe extends to scales greater than 400 Mpc \citep{gott05}.
\cite{clowes13} find a large quasar group that extends to
$500\,h^{-1}$ Mpc at $z \sim 1.3$.  More recently, \cite{lopez22}
report the discovery of an enormously large giant arc at $z\sim 0.8$
that spans $\sim 1$ Gpc. \cite{friday22} find correlated orientations
of the axes of large quasar groups on Gpc scales. A supervoid of
diameter $\sim 600 \, h^{-1}$ Mpc detected by \cite{keenan13},
indicates the possible existence of very large underdense regions in
the universe. Some other studies report the evidence of bulk flow from
the analysis of Type Ia supernovae \citep{colin19} and the existence
of anomalously large dipole in the distribution of quasars
\cite{secrest21}. These findings indicate the presence large-scale
inhomogeneity and anisotropy that are in apparent contradiction with
the cosmological principle. However, there can always be homogeneity
and isotropy on some larger scales. So it is difficult to falsify the
cosmological principle solely based on these observations. Further,
the statistical significance of these structures are questionable
\citep{nadathur, sheth11, park12}. In any case, these observations
are interesting in their own right and require further scrutiny to
arrive at a conclusion.

A wide variety of statistical tools are used to characterise the
inhomogeneities in the universe. The n-point hierarchy of the
correlation functions and their Fourier transforms, the polyspectra
\citep{peebles80} are widely used to study the inhomogeneous matter
distribution in cosmology. The Minkowski Functionals can measure the
morphology of the large-scale structures in the universe
\citep{mecke94}. The Kullback–Leibler Relative Information Entropy can
distinguish the local inhomogeneous mass density field from its
spatial average \citep{hosoya04, akerblom12}. \cite{wiegand10} show
that a non-negligible dynamical entanglement may arise due to the the
mutual information between spatially separated but causally connected
regions. \cite{czinner16} use the Tsallis relative entropy to
calculate the mutual information between spatially separated but
causally connected regions of the universe. \cite{shiba20} study the
Renyi mutual information between distant spatial regions in the vacuum
state of a free scalar field.  \cite{espinosa21} show that the mutual
information between two spatial regions may get enhanced due to
inflation. The mutual information of disjoint regions in higher
dimension is discussed in \cite{cardy13}.

In the present work, we want to calculate the Shannon mutual
information between the disjoint but causally connected regions in an
expanding universe. The spatial distributions of matter in any two
distant regions may have finite mutual information due to the presence
of large-scale structures and long-range correlations. We do not
consider an inhomogeneous universe for our current analysis. We
consider a homogeneous and isotropic universe and study the time
evolution of the mutual information between any two distant
regions. It would be interesting to investigate the role of different
constituents of the universe on the time evolution of the mutual
information between distant regions.

The plan of our work is as follows. We describe the time evolution of
the mutual information in an expanding universe in section 2, describe
the results in section 3 and present our conclusions in section 4.

\section{Mutual information and its time evolution}
The configuration entropy associated with the matter distribution over
a significantly large volume $V$ of the universe is defined as
\cite{pandey17},\\
\begin{eqnarray}
  S(t)=-\int \rho(\vec{r},t)\,\log \rho(\vec{r}, t)\,dV
\label{eq:sc}
\end{eqnarray}
The volume $V$ is subdivided into a number of subvolumes $dV$ and the
density $\rho(\vec{r},t)$ is measured within each of them. Here $r$
describes the comoving coordinate associated with the centre of the
subvolumes and $\rho(\vec{r},t)$ refers to the matter density in the
subvolumes. The density $\rho(\vec{r},t)$ is directly related with the
probability of finding a mass element within a subvolume.

Let us consider two large identical volumes $V$ separated by a large
distance. We label these two volumes as $A$ and $B$. The two regions
$A$ and $B$ are causally connected. The configuration entropy of the
two regions $A$ and $B$ can be written as,\\

\begin{eqnarray}
  S_{{\scaleto{A}{3.5pt}}}(t)=-\int
  \rho_{{\scaleto{A}{3.5pt}}}(\vec{r}, t)\,\log
  \rho_{{\scaleto{A}{3.5pt}}}(\vec{r}, t)\,dV_{{\scaleto{A}{3.5pt}}}
\label{eq:sca}
\end{eqnarray}
and
\begin{eqnarray}
  S_{{\scaleto{B}{3.5pt}}}(t)=-\int
  \rho_{{\scaleto{B}{3.5pt}}}(\vec{r}, t)\,\log
  \rho_{{\scaleto{B}{3.5pt}}}(\vec{r}, t)\,dV{{\scaleto{B}{3.5pt}}}
\label{eq:scb}
\end{eqnarray}

We consider $A$ and $B$ to be significantly large volumes so that the
universe can be treated as statistically homogeneous and isotropic on
those scales. We are only interested on scales where one can safely
use the linear perturbation theory to describe the evolution of the
configuration entropy in these regions.

\begin{figure}[h]
\centering
\includegraphics[width=13.8 cm]{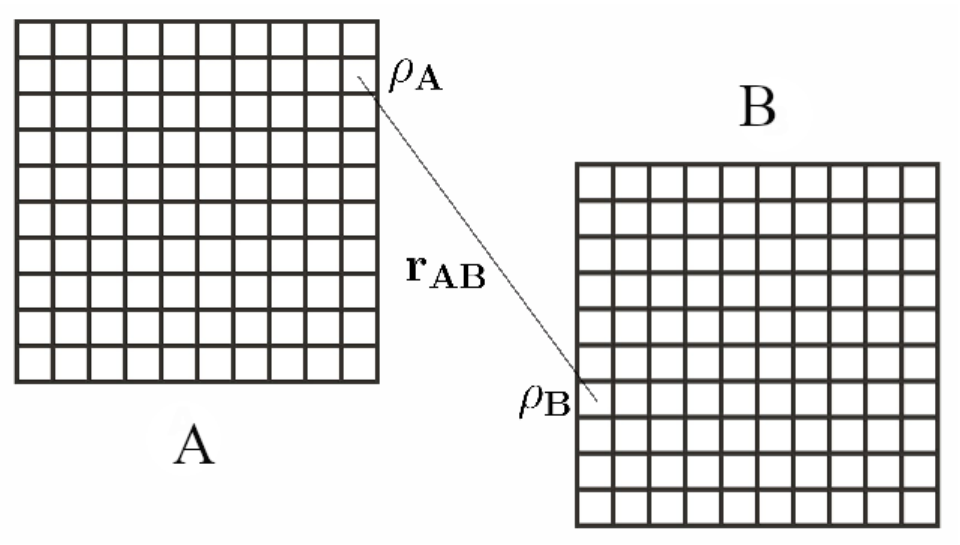}
\caption{This figure shows two large identical volumes $A$ and $B$
  divided into equal number of subvolumes. Here
  $\rho_{{\scaleto{A}{3.5pt}}}$ and $\rho_{{\scaleto{B}{3.5pt}}}$
  refers to the density within any two subvolumes at a given instant
  $t$ and $r_{{\scaleto{AB}{3.5pt}}}$ is the radial separation between
  the two subvolumes under consideration. We consider $A$ and $B$ to
  be causally connected.}\label{fig:fig1}
\end{figure} 

One can define the mutual information between the mass distributions
within the two regions $A$ and $B$ as,\\
\begin{eqnarray}
  I_{{\scaleto{AB}{3.5pt}}}(t)=\int \int
  \rho_{{\scaleto{AB}{3.5pt}}}(\vec{r_{1}}, t\,;\vec{r_{2}}, t)\,\log
  \frac{\rho_{{\scaleto{AB}{3.5pt}}}(\vec{r_{1}}, t\,;\vec{r_{2}},
    t)}{\rho_{{\scaleto{A}{3.5pt}}}(\vec{r_1},
    t)\,\rho_{{\scaleto{B}{3.5pt}}}(\vec{r_2},
    t)}\,dV_{{\scaleto{A}{3.5pt}}}\,dV_{{\scaleto{B}{3.5pt}}}
\label{eq:miab}
\end{eqnarray}
where $\rho_{{\scaleto{AB}{3.5pt}}}(\vec{r_{1}}, t\,;\vec{r_{2}}, t)$
is the joint density distributions in the two volumes. It provides the
joint probability of finding a mass element within each of the two
subvolumes. One can simplify \autoref{eq:miab} to write the mutual
information between the two regions as,\\
\begin{eqnarray}
  I_{{\scaleto{AB}{3.5pt}}}(t) = S_{{\scaleto{A}{3.5pt}}}(t) +
  S_{{\scaleto{B}{3.5pt}}}(t) - S_{{\scaleto{AB}{3.5pt}}}(t)
\label{eq:miabs}
\end{eqnarray}
where the joint entropy $S_{{\scaleto{AB}{3.5pt}}}(t)$ can be
expressed as,\\
\begin{eqnarray}
  \begin{split}
  S_{{\scaleto{AB}{3.5pt}}}(t)&=-\int \int
  \rho_{{\scaleto{AB}{3.5pt}}}(\vec{r_{1}}, t\,;\vec{r_{2}}, t)\,\log
  \rho_{{\scaleto{AB}{3.5pt}}}(\vec{r_{1}}, t\,;\vec{r_{2}},
  t)\,dV_{{\scaleto{A}{3.5pt}}}\,dV_{{\scaleto{B}{3.5pt}}}\\
  &=-\int \int
  \bar{\rho}_{{\scaleto{A}{3.5pt}}}(t)\,\bar{\rho}_{{\scaleto{B}{3.5pt}}}(t)\,[1+\xi(\vec{r}_{{\scaleto{AB}{3.5pt}}},t)]\,\log\Big[\bar{\rho}_{{\scaleto{A}{3.5pt}}}(t)\,\bar{\rho}_{{\scaleto{B}{3.5pt}}}(t)\,\left(1+\xi(\vec{r}_{{\scaleto{AB}{3.5pt}}},t)\right)\Big]\,dV_{{\scaleto{A}{3.5pt}}}\,dV_{{\scaleto{B}{3.5pt}}}
  \end{split}
\label{eq:scab}
\end{eqnarray}
where $\xi(\vec{r}_{{\scaleto{AB}{3.5pt}}},t)$ is the two-point
correlation function and $\bar{\rho}_{{\scaleto{A}{3.5pt}}}(t)$,
$\bar{\rho}_{{\scaleto{B}{3.5pt}}}(t)$ are the mean density in the two
regions $A$ and $B$. The separation vector between the centre of the
two subvolumes is
$\vec{r}_{{\scaleto{AB}{3.5pt}}}=\vec{r_{2}}-\vec{r_{1}}$. The
two-point correlation function
$\xi(\vec{r}_{{\scaleto{AB}{3.5pt}}},t)$ provides the excess
probability of finding two mass element separated by
$\vec{r}_{{\scaleto{AB}{3.5pt}}}$. The mutual information
$I_{{\scaleto{AB}{3.5pt}}}(t)$ quantifies the reduction of uncertainty
in the mass distribution within one volume given that we have the
complete knowledge of the mass distribution in the other. In other
words, it quantifies the expected gain in information about the mass
distribution in one volume when the other volume is observed.

We assume that the matter distribution in the universe is homogeneous
and isotropic on a scale of $V^{\frac{1}{3}}$. This allows us to write
$\bar{\rho}_{{\scaleto{A}{3.5pt}}}(t)\approx\bar{\rho}_{{\scaleto{B}{3.5pt}}}(t)
= \bar{\rho}(t)$, $S_{{\scaleto{A}{3.5pt}}}(t) \approx
S_{{\scaleto{B}{3.5pt}}}(t) = S(t)$,
$\xi(\vec{r}_{{\scaleto{AB}{3.5pt}}},t)=\xi(|\vec{r}_{{\scaleto{2}{3.5pt}}}-\vec{r}_{{\scaleto{1}{3.5pt}}}|,t)=\xi(r_{{\scaleto{AB}{3.5pt}}},t)$
and $ \frac{dS_{{\scaleto{A}{3.5pt}}}(t)}{dt} \approx
\frac{dS_{{\scaleto{B}{3.5pt}}}(t)}{dt} = \frac{dS(t)}{dt}$.

The mean density $\bar{\rho}(t)$ and the two-point correlation
function $\xi(r_{{\scaleto{AB}{3.5pt}}},t)$ would evolve differently
in different cosmological models. The mutual information between the
two regions and its time evolution would thus depend on the
cosmological model.

\section{Results}
\subsection{Mutual information in a matter dominated Universe}
We would like to calculate the mutual information between the mass
distributions in the two regions $A$ and $B$ in a matter dominated
Universe ($\Omega_m=1$). The average density in a matter dominated
Universe is $\bar{\rho}(t)=\frac{1}{6\,\pi\,G\,t^2}$ and the growing
mode of density perturbations is $D(t) \propto t^{\frac{2}{3}}$. In
the linear regime, the time evolution of the two-point correlation
function can be described as,
$\xi(r_{{\scaleto{AB}{3.5pt}}},t)=D^2(t)\,\xi(r_{{\scaleto{AB}{3.5pt}}})$. Here
the proportionality constant is absorbed in
$\xi(r_{{\scaleto{AB}{3.5pt}}})$.

\cite{pandey17} shows that the configuration entropy rate in a matter
dominated Universe is always negative i.e. $ \frac{dS(t)}{dt}<0$. Let
us write $ \frac{dS(t)}{dt}=-f(t)$.

The time evolution of the mutual information between the regions $A$
and $B$ can be written from \autoref{eq:miabs} as,
\begin{eqnarray}
  \frac{d I_{{\scaleto{AB}{3.5pt}}}(t)}{dt} = -2 \frac{dS(t)}{dt}+\frac{d}{dt}
  \int \int
  \Big[\bar{\rho}^2(t)\,\left(1+D^{2}(t)\,\xi(r_{{\scaleto{AB}{3.5pt}}})\right)\Big]
  \,\log
  \Big[\bar{\rho}^2(t)\,\left(1+D^{2}(t)\,\xi(r_{{\scaleto{AB}{3.5pt}}})\right)\Big]
  \,dV_{{\scaleto{A}{3.5pt}}}\,dV_{{\scaleto{B}{3.5pt}}}
\label{eq:miedu}
\end{eqnarray}

Simplifying \autoref{eq:miedu}, we get,

\begin{eqnarray}
 \frac{d I_{{\scaleto{AB}{3.5pt}}}(t)}{dt} =-2 f(t)+I_{1}+I_{2}+I_{3}+I_{4}+I_{5}
\label{eq:miedu1}
\end{eqnarray}

where 

\begin{eqnarray}
  I_{1} =\frac{2}{9\,\pi^{2}\,G^{2}}\,\log(6\pi G) \int \int
  \left(t^{-5}+\frac{2}{3}\,t^{-\frac{11}{3}}
  \,\xi(r_{{\scaleto{AB}{3.5pt}}})\right)\,dV_{{\scaleto{A}{3.5pt}}}\,dV_{{\scaleto{B}{3.5pt}}}
\label{eq:i1}
\end{eqnarray}

\begin{eqnarray}
  I_{2} =-\frac{1}{9\pi^{2}G^{2}}\, \int \int t^{-5} \left(1-4\log
  t\right)\,dV_{{\scaleto{A}{3.5pt}}}\,dV_{{\scaleto{B}{3.5pt}}}
\label{eq:i2}
\end{eqnarray}

\begin{eqnarray}
  I_{3} = -\frac{1}{9\pi^{2}G^{2}}\, \int \int
  \xi(r_{{\scaleto{AB}{3.5pt}}})\,t^{-\frac{11}{3}}
  \left(1-\frac{8}{3}\,\log
  t\right)\,dV_{{\scaleto{A}{3.5pt}}}\,dV_{{\scaleto{B}{3.5pt}}}
\label{eq:i3}
\end{eqnarray}

\begin{eqnarray}
  I_{4} =-\frac{1}{9\pi^{2}G^{2}}\, \int \int
  \left(t^{-5}+\frac{2}{3}\,t^{-\frac{11}{3}}\,\xi(r_{{\scaleto{AB}{3.5pt}}})\right)\,\log
  \left(1+t^{\frac{4}{3}}\,\xi(r_{{\scaleto{AB}{3.5pt}}})\right)\,dV_{{\scaleto{A}{3.5pt}}}\,dV_{{\scaleto{B}{3.5pt}}}
\label{eq:i4}
\end{eqnarray}
and
\begin{eqnarray}
  I_{5} = \frac{1}{27\pi^{2}G^{2}}\, \int \int
  \left(t^{-4}+t^{-\frac{8}{3}}\,\xi(r_{{\scaleto{AB}{3.5pt}}})\right)\,\frac{t^{\frac{1}{3}}\,\xi(r_{{\scaleto{AB}{3.5pt}}})}{1+t^\frac{4}{3}\,\xi(r_{{\scaleto{AB}{3.5pt}}})}\,dV_{{\scaleto{A}{3.5pt}}}\,dV_{{\scaleto{B}{3.5pt}}}
\label{eq:i5}
\end{eqnarray}

The integrals in the expressions of $I_{1}, I_{2}, I_{3}, I_{4}$ and
$I_{5}$ can not be simplified further due to the lack of
symmetry. However, one can easily analyze the time dependence of these
expressions. Observations show that the galaxy two-point correlation
function has a nearly universal dependence on pair separation $r$ as
$\xi(r) \sim r^{-1.8}$. The terms involving
$\xi(r_{{\scaleto{AB}{3.5pt}}})$ in these expressions will have a
smaller magnitude. The \autoref{eq:i1}, \autoref{eq:i2},
\autoref{eq:i3}, \autoref{eq:i4} and \autoref{eq:i5} have strong time
dependence. $I_{2}$ and $I_{3}$ become positive for larger values of
time. Only $I_{4}$ remains negative at all time. The sum
$I=(I_{1}+I_{2}+I_{3}+I_{4}+I_{5})$ is positive but decays towards
zero with increasing time. The term $f(t)$ in \autoref{eq:miedu1} has
a much weaker time dependence as compared to $I$
\citep{pandey17}. This leads to $\frac{d
  I_{{\scaleto{AB}{3.5pt}}}(t)}{dt}<0$ which implies that the mutual
information between two independent regions decreases with time in a
matter dominated Universe.

It may be noted that \autoref{eq:miedu1} depends on the size of the
regions $A$ and $B$ and the separation between them.  The integrals in
\autoref{eq:miedu1} would be carried out over different volumes when
there is a change in the size of the two regions. The separations
between the different pairs of subvolumes would change with the
distance between the two regions. The integrals in \autoref{eq:miedu1}
will have different values since the two-point correlation function
changes with the separation.  If the two regions are separated by a
very large distance compared to the dimensions of the two regions then
the integrals in \autoref{eq:miedu1} would loss their physical
relevance.

\subsection{Mutual information in a $\Lambda$-dominated Universe}
Now we would like to calculate $\frac{d
  I_{{\scaleto{AB}{3.5pt}}}(t)}{dt}$ in a $\Lambda$-dominated
Universe. We have $\bar{\rho}(t)=\bar{\rho}=$constant and the growing
mode of density perturbations is $D(t)=k=$ constant in
$\Omega_{\Lambda}=1$ Universe. The time evolution of the mutual
information in such a universe can be expressed as,\\
\begin{eqnarray}
  \frac{d I_{{\scaleto{AB}{3.5pt}}}(t)}{dt} = -2
  \frac{dS(t)}{dt}+\frac{d}{dt} \int \int
  \Big[\bar{\rho}^2\,\left(1+k^{2}\,\xi(r_{{\scaleto{AB}{3.5pt}}})\right)\Big]
  \,\log
  \Big[\bar{\rho}^2\,\left(1+k^{2}\,\xi(r_{{\scaleto{AB}{3.5pt}}})\right)\Big]
  \,dV_{{\scaleto{A}{3.5pt}}}\,dV_{{\scaleto{B}{3.5pt}}}
\label{eq:milam}
\end{eqnarray}

\citep{pandey17} show that $\frac{dS}{dt}=0$ in a $\Lambda$-dominated
Universe. So we have $ \frac{d I_{{\scaleto{AB}{3.5pt}}}(t)}{dt}=0$.
Clearly, $I_{{\scaleto{AB}{3.5pt}}}=$ constant in a
$\Lambda$-dominated Universe. There would be a constant mutual
information between the regions $A$ and $B$ at all times in such a
universe.

\subsection{Mutual information in the $\Lambda$CDM model, the dynamical dark energy models and the holographic dark energy models}
It is clear that the joint entropy between the two regions $A$ and $B$
plays a negligible role in the time evolution of the mutual
information. The configuration entropy rate $\frac{dS}{dt}$ determines
the time evolution of the mutual information between the two
regions. The configuration entropy rates have been calculated for the
$\Lambda$CDM model, different dynamical dark energy models and the
holographic dark energy models in the literature \citep{das19,
  pandey19a, bhattacharjee21, das23}. The configuration entropy rate
decreases to reach a minimum and then increases with time in all these
models. However, the location and the amplitude of the minimum depend
on the models. The location of the minimum precisely indicates the
epoch of dark energy domination predicted by the relevant model. The
average density will fall faster in such models as compared to the
matter dominated Universe. So the joint entropy term would contribute
negligibly to the evolution of the mutual information in all these
models. The time evolution of the mutual information in a given model
will be thus entirely determined by the behaviour of the configuration
entropy rate in that model. The mutual information between $A$ and $B$
in all these models would initially decrease with time and eventually
hit a minimum. The mutual information would increase after this
minimum once the dark energy starts to dominate the dynamics of the
universe.

\section{Conclusions}
We analyze the time evolution of the mutual information between
disjoint regions of the universe in different cosmological models. The
mutual information here quantifies the reduction of uncertainty in the
knowledge of the matter distribution in one region given that we have
complete knowledge of it in the other region. In another words, the
mutual information provides some knowledge about the matter
distribution in one region provided we have the complete knowledge of
the matter distribution in the another region. A zero mutual
information indicates that the mass distribution in the two regions
are statistically independent. We do not separately calculate the
mutual information between the two disjoint regions but obtain an
expression for its time evolution from the definition. It may be noted
that the mutual information is positive or zero by definition. It can
not assume negative values. However, the rate of change of the mutual
information can be negative, positive or zero.

The time evolution of the mutual information between disjoint regions
of the universe is primarily determined by the dynamics of the
expansion and the growth rate of the density perturbations. We find
that the mutual information decreases continuously in a purely matter
dominated Universe whereas it stays constant in a purely
$\Lambda$-dominated Universe. So the disjoint regions become
statistically independent in a matter-dominated Universe whereas they
remain entangled forever in a $\Lambda$-dominated Universe. The mutual
information decreases to reach a minimum and then increases with time
in the $\Lambda$CDM model, dynamical dark energy models and the
holographic dark energy models.  Clearly, the time evolution of the
mutual information is governed by the changes in the configuration
entropy of the matter distribution in the universe. The change in the
joint entropy between the mass distributions in the two regions does
not contribute significantly to this evolution.

The two regions $A$ and $B$ are causally connected. However, they may
be separated by large distance. In reality, one can not measure the
mass distributions in the two volumes simultaneously. One can infer
some information about the mass distribution in one volume while
observing the other. These information corresponds to the same cosmic
time. It is worthwhile to mention here that a mutual information
between the two causally connected regions may also introduce a
non-negligible dynamical entanglement \citep{wiegand10}. The effect of
such dynamical entanglement is not considered in the present work.
Further, one can also consider the contributions from the higher order
correlations. A non-zero three-point correlation function would modify
the joint probabilities in \autoref{eq:miab}. We plan to address these
issues in future works. It would be also interesting to study the
evolution of mutual information in the inhomogeneous cosmological
models. The presence of long range correlations in the mass
distributions can significantly modify the mutual information and its
evolution.

The continuous dissipation of the configuration entropy during the
matter dominated era demands enormous entropy production that can
counterbalance this loss and maximize the entropy \citep{pandey17}.
The accelerated expansion of the universe provides an avenue for
maximum entropy production in accordance with the second law of
thermodynamics. It is interesting to note that the evolution of the
mutual information is strongly sensitive to the cosmological constant
or dark energy. This implies that the mutual information may have
deeper connections to the dark energy and the accelerated expansion of
the universe.

\vspace{6pt} 

\section{Acknowledgements}
The author thanks the anonymous reviewers for some valuable comments and suggestions on the dtaft. This research was funded by the SERB, DST, Government of India through the project CRG/2019/001110. The author would also like to acknowledge IUCAA, Pune for providing support through associateship programme.



%


\end{document}